\RequirePackage{fix-cm}
\documentclass[smallcondensed]{svjour3}     
\smartqed  
\usepackage{graphicx}
\usepackage[utf8x]{inputenc}

\usepackage{subcaption}
\usepackage{wrapfig}

\usepackage{amsmath}
\usepackage{amssymb}
\usepackage{upgreek}
\usepackage{color}
\definecolor{keywordcolor}{rgb}{0.7, 0.1, 0.1}   
\definecolor{tacticcolor}{rgb}{0.0, 0.1, 0.3}    
\definecolor{commentcolor}{rgb}{0.4, 0.4, 0.4}   
\definecolor{stringcolor}{rgb}{0.5, 0.3, 0.2}    
\definecolor{symbolcolor}{rgb}{0.1, 0.2, 0.7}    
\definecolor{sortcolor}{rgb}{0.1, 0.5, 0.1}      
\definecolor{attributecolor}{rgb}{0.7, 0.1, 0.1} 
\definecolor{errorcolor}{rgb}{1, 0, 0}           

\usepackage{listings}
\usepackage{scalefnt}
\usepackage{relsize}
\usepackage{hyperref}

\hypersetup{
    colorlinks=true,
    citecolor=blue,
    linkcolor=blue,
    urlcolor=blue,
}
\usepackage{verbatim}

\newlength\myverbindent
\makeatletter
\newcommand{\verbatimfont}[1]{\def\verbatim@font{#1}}%
\def\verbatim@processline{%
  \hspace{\myverbindent}\the\verbatim@line\par}
\makeatother
\verbatimfont{\fontfamily{qcr}\selectfont\scalefont{.93}}

\newcommand{\mm}[1]{\mbox{\fontfamily{qcr}\selectfont\scalefont{.93}#1}}
\newcommand{\lean}[1]{\lstinline[language=lean]{#1}}
\usepackage{url}
\usepackage{xspace}
\newcommand{\mathlib}{\textsf{mathlib}\xspace}

\journalname{Preprint}

\begin{document}

\title{A bi-directional extensible interface \\ between Lean and Mathematica
\footnote{
    This paper expands on a workshop paper by the first author~\cite{lewis:17},
    which describes an early version of one direction of the interface.}
\thanks{
The first author receives support
from the European Research Council (ERC) under
the European Union's Horizon 2020 research and innovation
program (grant agreement No. 713999, Matryoshka)
and from the Dutch Research Council (NWO) under the
Vidi program (project No. 016.Vidi.189.037, Lean Forward).
}
}


\author{Robert Y. Lewis         \and
        Minchao Wu 
}


\institute{Robert Y. Lewis \at
              Vrije Universiteit Amsterdam \\
              \email{r.y.lewis@vu.nl}           
           \and
           Minchao Wu \at
              Australian National University \\
              \email{logic.mcwu@gmail.com}
}

\date{Received: date / Accepted: date}

\maketitle

\begin{abstract}

We implement a user-extensible ad hoc connection
between the Lean proof assistant
and the computer algebra system Mathematica.
By reflecting the syntax of each system in the other
and providing a flexible interface for extending translation,
our connection allows for
the exchange of arbitrary information between the two systems.

We show how to make use of the Lean metaprogramming framework
to verify certain Mathematica computations,
so that the rigor of the proof assistant is not compromised.
We also use Mathematica as an untrusted oracle
to guide proof search in the proof assistant
and interact with a Mathematica notebook
from within a Lean session.
In the other direction,
we import and process Lean declarations from within Mathematica.
The proof assistant library serves as a database of mathematical knowledge
that the CAS can display and explore.

\end{abstract}
\section{Introduction}
\label{section:introduction}

Many researchers have noted the disconnect
between computer algebra and interactive theorem proving.
In the former, one typically values speed and flexibility over absolute correctness.
To be more efficient or user-friendly,
a computer algebra system (CAS) may
blur the distinction between polynomial objects and polynomial functions,
assume that sufficiently small terms at the end of a series are zero,
or resort to numerical approximations without warning.
Such simplifying assumptions often make sense in the context of computer algebra;
the capability and flexibility of these systems
make them indispensable tools to many working mathematicians.
These assumptions, though,
are antithetical to the goals of interactive theorem proving (ITP),
where every inference must be justified by appeal to some logical principle.
The strict logical requirements
and lack of familiar algorithms
discourage many mathematicians from using proof assistants.
Conversely, the unreliability of many computer algebra systems,
and their lack of proof languages and proof libraries,
often makes them unsuitable for mathematical justification.

Integrating computer algebra and proof assistants
is one way to reduce this barrier to entry to ITP
and to strengthen the justificatory power of computer algebra.
Bridges between the two types of systems have been built in a variety of ways.
We contribute another such bridge,
between the proof assistant Lean \cite{demoura:14}
and the computer algebra system Mathematica \cite{Wolfram2015}.
Since Mathematica is one of the most commonly used computer algebra systems,
and a user with knowledge of the CAS can extend the capabilities of our link,
we hope that the familiarity will lead to wider use.
Our connection is inspired by the architecture described by Harrison and Th\'ery \cite{Harrison1998}.

A number of features of our bridge distinguish it from earlier work.
CAS results imported into the proof assistant can be trusted, verified, or used ephemerally
to guide proof development.
The translation can be extended in-line with library development
without modifying auxiliary dictionaries or source code.
We are able to simulate a Mathematica read--eval--print loop (REPL)
in a Lean editor session,
which takes as input Mathematica expressions
with arbitrary Lean expressions inserted,
evaluates these in Mathematica,
and displays the result.
The link works bi-directionally using the same translation procedure,
allowing Mathematica to access Lean's library and automation.
The link requires no plugins or modified source code:
it is available as a standard Lean library.



Our link separates the steps of communication, semantic interpretation, and verification:
there is no a priori restriction on the type of information
that can be shared between the systems.
With the proof assistant in the leading role,
Lean expressions are exported to Mathematica,
where they can be interpreted and manipulated.
The results are then imported back into Lean and reinterpreted.
One can then write scripts that verify properties of the translated results.
This style of interaction,
where verification happens on a per-case basis after the computation has ended,
is called \emph{ad hoc}.

By performing calculations in Mathematica and verifying the results in Lean,
we relax neither the rigor of the proof assistant
nor the efficiency of the CAS.
Alternatively, we can trust the CAS as an oracle,
or use it in a purely informative role,
where its output does not appear in the final proof term.
We provide comprehensive tactics to perform and verify certain classes of computations,
such as factoring polynomials and matrices.
But all the components of our procedure are implemented transparently
in Lean's metaprogramming framework,
and they can easily be extended or used for one-off computations from within the Lean environment.

This range of possibilities is intended
to make our link attractive to multiple audiences.
The working mathematician or mathematics student,
who balks at the restrictions imposed by a proof assistant,
may find that full access to a familiar CAS is worth the tradeoff in trust.
Industrial users are often happy to trust
both large-kernel proof assistants and computer algebra systems,
and the rigor of Lean with Mathematica as an oracle falls somewhere in between.
Certifiable algorithms are still available to users who demand complete trust.
The ease of metaprogramming in Lean is another draw:
users do not need to learn a new programming or tactic language
to write complicated translation rules or verification procedures.

The translation procedure used is symmetric
and can be used for communication in the reverse direction as well.
Mathematica has no built-in notion of proof,
although it does have head symbols that express propositions.
Rather than establishing an entire proof calculus
for these symbols within Mathematica,
we export theorem statements to Lean,
where they can be verified in an environment designed for this purpose.
The resulting proof terms are interpreted in the CAS
and can be displayed or processed as needed.
Alternatively, we can skip the verification step
and display lemmas that are likely to be relevant to Mathematica's goal.
In some sense,
the link allows Mathematica to ``borrow''
Lean's semantics, proof language, and proof library.


The source for this project, and supplementary documentation,
is available online.\footnote{\url{https://robertylewis.com/leanmm/}}
In this paper, we use \lean{Computer Modern} for Lean code
and \mm{TeX} \mm{Gyre} \mm{Cursor} for Mathematica code.


\section{System descriptions}
\label{section:systems}

\subsection{Lean}
\label{subsection:systems:lean}
Lean is a proof assistant developed at Microsoft Research \cite{demoura:14}.
It is based on the Calculus of Inductive Constructions (CIC) \cite{Coquand1988,Coquand1990},
an extension of the lambda calculus with dependent types and inductive definitions.
There is a non-cumulative hierarchy of type universes \lean{Sort u}, \lean{u ≥ 0},
with abbreviations \lean{Prop = Sort 0} and \lean{Type u = Sort (u+1)}.
The bottom level \lean{Prop} is impredicative and proof-irrelevant.
The Lean community develops \mathlib,
a rapidly growing library of verified mathematics, programming, and tactics;
more details on Lean and its library can be found in the \mathlib system description~\cite{mathlib20}.

Lean's standard library uses type classes to implement an abstract algebraic hierarchy.
Arithmetic operations, such as $+$ and $*$,
and numerals are generic over types that instantiate the appropriate classes.
As an example, consider the signature of the addition operator:
\begin{lstlisting}[language=lean]
add.{u} : Π {A : Type u} [has_add A], A → A → A
\end{lstlisting}
The notation \lean{\{A : Type u\}} signals that the argument \lean{A} is an implicit variable, meant to be inferred from further arguments; \lean{has_add : Type u → Type u} is a type class, and the notation \lean{[has_add A]} signals that a term of that type is to be inferred using type class resolution. The universe argument \lean{u} indicates that \lean{add} is parametric over one universe level.

The dependently typed language implemented in Lean
is flexible enough to serve as its own \emph{metaprogramming language} \cite{demoura:17}.
Data types and procedures implemented in Lean's C++ code base
are exposed as constants,
using the keyword \lean{meta}
to mark a distinction between the object language and this extension.
Expressions can be evaluated in the Lean virtual machine,
which replaces these constants with their underlying implementations.
Meta-definitions permit unbounded recursion
but are otherwise quite similar to standard definitions.

Combined with the declaration of the types \lean{expr} and \lean{pexpr},
which expose the syntax of elaborated and unelaborated Lean expressions in Lean itself,
and \lean{tactic_state},
which exposes the environment and goals of a tactic proof,
this metaprogramming framework allows users
to write complex procedures for constructing proofs.
A term of type \lean{tactic A} is a function
\begin{lstlisting}[language=lean]
tactic_state → tactic_result A
\end{lstlisting}
where a result is either a success
(pairing a new \lean{tactic_state} with a term of type \lean{A})
or a failure.
Proof obligations can be discharged by terms of type \lean{tactic unit}.
Such a term is executed in the Lean virtual machine
to transform the original \lean{tactic_state}
into one in which all goals have been solved.
More generally,
we can think of a term of type \lean{tactic A}
as a program that attempts to construct a term of type \lean{A},
while optionally changing the tactic state.

When writing tactics, the command \lean{do} enables Haskell-like monadic syntax. For example, the following tactic returns the number of goals in the current tactic state.
The type of \lean{get_goals} is \lean{tactic (list expr)}, where \lean{list} is the standard (object-level) type defined in the Lean library.
\begin{lstlisting}[language=lean]
meta def num_goals : tactic nat :=
do gs ← get_goals,
   return (length gs)
\end{lstlisting}

Lean allows the user to tag declarations with \emph{attributes}, and provides an interface \lean{name → tactic (list name)}
to retrieve a list of declarations tagged with a certain attribute.

%

\subsection{Mathematica}
\label{subsection:systems:mathematica}
Mathematica is a popular symbolic computation system developed at Wolfram Research,
implementing the Wolfram Language \cite{Wolfram2015}.
Along with support for a vast range of mathematical computations,
Mathematica includes collections of data of various types and tools for manipulating this data.

Mathematica provides comprehensive tools for rewriting and solving polynomial, trigonometric, and other classes of equations and inequalities; solving differential equations, both symbolically and numerically; computing derivatives and integrals of various types; manipulating matrices; performing statistical calculations, including fitting and hypothesis testing; and reasoning with classes of special functions.

This large library of functions is one reason to choose Mathematica for our linked CAS.
Another reason is its ubiquity:
Mathematica is frequently used in undergraduate mathematics and engineering curricula.
Lean beginners who are accustomed to Mathematica
do not need to learn a new CAS language for the advanced features of this link.
The Wolfram language is a symbolic functional language with a simple grammar,
making it a good candidate for intertranslation with Lean
without having to represent low-level data structures.

For those unfamiliar with the syntax of the Wolfram Language, we note some features and terminology that will help to understand the code fragments in this paper.
\begin{itemize}
 \item Function application is written using square brackets, e.g. \mm{Plus[x, y]}.
  Many functions are variadic:
  that is, one can also write \mm{Plus[x, y, z]}.
  Common notation like \mm{x + y + z} is also supported.
 \item Alternatively, one can write unary function application in postfix form: \\
  \mm{x\^{}2 - 2x + 1 // Factor} is equivalent to \mm{Factor[x\^{}2 - 2x + 1]}.
 \item In the expression \mm{Plus[x, y]},
   we refer to \mm{Plus} as the \emph{head symbol}
   and \mm{x} and \mm{y} as the \emph{arguments}.
   Non-numeric atoms like \mm{Plus}, \mm{x}, and \mm{y} are called \emph{symbols}.
 \item There is no strong distinction between defined and undefined symbols.
   The user is free to introduce a new symbol and use it at will.
   The computational behavior of a head symbol can be fully or partially defined via pattern matching rules,
   such as \mm{F[x\_,y\_] := x+y}.
   The underscores indicate that \mm{x\_} and \mm{y\_} are patterns.
 \item The Wolfram Language is untyped,
   so head symbols such as \mm{Plus} and \mm{Factor}
   can be applied to any argument or sequence of arguments.
   Evaluation is often restricted to certain patterns.
   For example, \mm{Plus[2, 3]} will evaluate to \mm{5}
   but \mm{Plus[Factor, Plus]} will not reduce.
   Nevertheless, both are well-formed Mathematica expressions.
\end{itemize}
\vspace{-1em}

\section{The translation procedure}
\label{section:translation}
Our bridge can be used to exchange information between Mathematica and Lean.
The logical foundations and semantics of the two systems are quite different,
and we should not expect a perfect correspondence between the two.
However, in many situations,
an expression in Lean has a counterpart in Mathematica
with a very similar intended meaning.
We can exploit these similarities
by ignoring the unsoundness of the translations in both directions
and attempting to verify, post hoc, that the resulting expression has the intended properties.

As a running example, suppose we want to show in Lean:
\begin{lstlisting}[language=lean]
x : real ⊢ x^2 - 2x + 1 ≥ 0
\end{lstlisting}
Factoring the left-hand side of the inequality makes this a one-step proof (assuming we've proven that squares are nonnegative). It is nontrivial to write a reliable and efficient polynomial factoring algorithm, but luckily, one is implemented in Mathematica. So we would like to do the following:
\begin{enumerate}
\item Transform the Lean representation of $x^2 - 2x + 1$ into Mathematica syntax.
\item Interpret this as the Mathematica representation of the same polynomial.
\item Use Mathematica's \mm{Factor} function to factor the polynomial.
\item Transform this back into Lean syntax, and interpret it as a Lean polynomial.
\item Verify that the new expression is equal to the old.
\item Substitute this equality into the goal.
\end{enumerate}

In Section \ref{subsection:translation:lm} we describe steps 1, 2, and 4. Once we have a valid Mathematica expression, step 3 is trivial. We discuss steps 5 and 6 in Section \ref{section:mmfromlean}; since checking that a polynomial has been factored correctly is much easier than factoring it in the first place, these are handled easily by simplification and rewriting.

It is worth emphasizing the modularity and extensibility of this approach. Both directions of translation are handled independently, and the translation rules can be extended or changed at will. Translation rules may be arbitrarily complex. Users may choose to use alternate verification procedures, or to forgo the verification step entirely.

\subsection{Translating Lean to Mathematica}
\label{subsection:translation:lm}

The Lean expression grammar is presented (in Lean syntax) in Figure~\ref{figure:leankinds}.
The type \lean{expr} is marked with the keyword \lean{meta}
because, during evaluation,
the Lean virtual machine replaces terms of this type with the kernel's expression datatype.
In the explanation below, we focus on the parts of interest for our link.
In particular we will not discuss the \lean{macro} constructor.

\begin{figure}[t!]
\begin{lstlisting}[language=lean]
meta inductive expr
| var         : nat → expr
| sort        : level → expr
| const       : name → list level → expr
| mvar        : name → expr → expr
| local_const : name → name → binder_info → expr → expr
| app         : expr → expr → expr
| lam         : name → binder_info → expr → expr → expr
| pi          : name → binder_info → expr → expr → expr
| elet        : name → expr → expr → expr → expr
| macro       : macro_def → list expr → expr
\end{lstlisting}
\caption{
    The Lean expression grammar is captured by the type \lean{expr}.
    Every Lean expression is uniquely expressed
    using one of these constructors.
}
\label{figure:leankinds}
\end{figure}

Each Lean expression exists in an environment, which contains the names, types, and definitions of previous declarations. The \lean{const} kind accesses a previous declaration, instantiated to particular universe levels if the declaration is parametric. In addition to declarations in its environment, an expression may refer to its local context, which contains variables and hypotheses of kind \lean{local_const}. In the toy example introduced above, \lean{x} is a local constant. A local constant has a unique name, a formatting name, and a type.

The expression kinds \lean{lam} and \lean{pi} respectively represent lambda-abstraction and the dependent function type. (Non-dependent function types are degenerate cases of pi types.) Each contains a name for the bound variable, the type of the variable, and the expression body. Bound variables of kind \lean{var} are anonymous within the body, being represented by De Bruijn indices \cite{McBride2004}. Application of one expression to another is represented by the \lean{app} kind.

Type universes are implemented by the expression kind \lean{sort}. Metavariables represent placeholders in partially constructed expressions; the \lean{mvar} kind holds the name and type of the placeholder. Let expressions (\lean{elet}) bind a named variable with a type and value within a body.

To represent this syntax in Mathematica, we define
\begin{lstlisting}[language=lean]
mathematica_form_of_expr : expr → string
\end{lstlisting}
by recursion over the \lean{expr} datatype. We associate a Mathematica head symbol \mm{LeanVar}, \mm{LeanSort}, \mm{LeanConst}, etc.\ to each constructor of \lean{expr}. Names, levels, lists of levels, and binder information are also represented.

Some of the information contained in a Lean expression has little plausible use in Mathematica, or is needlessly verbose: for example, it is hard to contrive a scenario in which the full structure of a Lean \lean{name} is used in the CAS. Nonetheless, we do not strip any information at this stage, to preserve that an expression reflected into and immediately back from Mathematica should translate to the original expression without having to inject any additional information.

In our running example, we work on the expression $x^2 - 2x + 1$.
The fully elaborated Lean expression and its Mathematica representation are too long to print here,
but they can be viewed in the supplementary documentation.
Instead, we consider the more concise example of $x+x$.
If we use strings to stand in for terms of type \lean{name},
natural numbers in place of universe levels,
and the string \lean{"bi"} in place of the default \lean{binder_info} argument,
and we abbreviate
\begin{lstlisting}[language=lean]
𝒳 := local_const "x" "x" "bi" (const "real" []),
\end{lstlisting}
we can write the full form of $x+x$:
\begin{lstlisting}[language=lean]
app (app (app (app (const "add" [0]) (const "real" []))
              (const "real.has_add" [])) 𝒳) 𝒳
\end{lstlisting}
The corresponding Mathematica expressions are
\begin{verbatim}
X := LeanLocal["x", "x", "bi", LeanConst["real", {}]]

LeanApp[LeanApp[LeanApp[LeanApp[LeanConst["add", {0}],
                                LeanConst["real", {}]],
                        LeanConst["real.has_add", {}]], X], X]
\end{verbatim}

Since the head symbols \mm{LeanApp}, \mm{LeanConst}, etc.\ are uninterpreted in Mathematica, this representation is not yet useful. We wish to exploit the fact that many Lean terms have semantically similar counterparts in Mathematica. For instance, the Lean constants \lean{add} and \lean{mul} behave similarly to the Mathematica head symbols \mm{Plus} and \mm{Times}; both systems have notions of application, although they handle the arity of applications differently; and Mathematica's concept of a ``pure function" is analogous to lambda-abstraction in Lean.

We thus define a translation function \mm{LeanForm} in Mathematica that attempts to interpret the syntactic representation. Mathematica functions are typically defined using pattern matching. The \mm{LeanForm} function, then, will look for familiar patterns (e.g. \lean{add A h x y}, in Mathematica syntax) and rewrite them in translated form (e.g. \mm{Plus[LeanForm[x],} \mm{LeanForm[y]]}). Users can easily extend this translation function by asserting additional equations; a default collection of equations is loaded automatically.

For our factorization example, we want to convert Lean arithmetic to Mathematica arithmetic. Among other similar rules, we will need the following:
\begin{verbatim}
LeanForm[LeanApp[LeanApp[LeanApp[LeanApp[
  LeanConst["add",_], _], _], x_], y_]] :=
Inactive[Plus][LeanForm[x],LeanForm[y]]
\end{verbatim}

Note that this pattern ignores the type argument and type-class instance in the Lean term. These arguments are irrelevant to Mathematica and can be inferred again by Lean in the back-translation.
We block Mathematica's computation with the \mm{Inactive} head symbol; otherwise, Mathematica would eagerly simplify the translated expression, which can be undesirable. The function \mm{Activate} strips these annotations, allowing reduction.

Numerals in Lean are type-parametric and are represented using the constants \lean{zero}, \lean{one}, \lean{bit0}, and \lean{bit1}. To illustrate, the type signature of the latter is
\begin{lstlisting}[language=lean]
bit1.{u} : Π {A : Type u}, [has_add A] → [has_one A] → A → A
\end{lstlisting}
and the numeral 6 is represented as \lean{bit0 (bit1 one)}.
The type of this numeral is expected to be inferable from context.
We can use rules similar to the above to transform Lean numerals into Mathematica integers:
\begin{verbatim}
LeanForm[LeanApp[LeanApp[LeanApp[LeanApp[
  LeanConst["bit1", _], _], _], _], t_]] :=
2*LeanForm[t]+1
\end{verbatim}

Applying \mm{LeanForm} will not necessarily remove all occurrences
of the head symbols \mm{LeanApp}, \mm{LeanConst}, etc.
This is not a problem:
we only need to translate the ``concepts'' with equivalents in Mathematica.
Unconverted subterms---for instance \mm{X},
which contains applications of \mm{LeanLocal} and \mm{LeanConst}---will
be treated as uninterpreted constants by Mathematica,
and the back-translation described below will return them to their original Lean form.

In our running example (keeping the abbreviation \mm{X}),
applying the \mm{LeanForm} and \mm{Activate} functions produces:
\begin{verbatim}
Plus[1,Times[-2, X], Power[X, 2]]
\end{verbatim}
Applying \mm{Factor} produces \mm{Power[Plus[-1, X], 2]}.


\subsection{Translating Mathematica to Lean}
\label{subsection:translation:ml}

Mathematica expressions are composed of various atomic number types, strings, symbols, and applications, where one expression is applied to a list of expressions. We represent this structure in Lean with the data type \lean{mmexpr} (Figure~\ref{figure:mmexpr}).
\begin{figure}[!t]
\begin{lstlisting}[language=lean]
inductive mmexpr
| sym   : string → mmexpr
| mstr  : string → mmexpr
| mint  : int → mmexpr
| app   : mmexpr → list mmexpr → mmexpr
| mreal : float → mmexpr
\end{lstlisting}
\caption{
    The Mathematica expression grammar is captured by the type \lean{mmexpr}.
    Every Mathematica expression is uniquely expressed
    using one of these constructors.
}
\label{figure:mmexpr}
\end{figure}

The result of a Mathematica computation is reflected into Lean as a term of type \lean{mmexpr}.
This is analogous to the original export of our Lean expression into Mathematica.
It remains to interpret it as something meaningful.

A \emph{pre-expression} in Lean is a term where universe levels and implicit arguments are omitted.
It is not expected to type-check,
but one can try to convert it into a type-correct term via elaboration.
For instance, the pre-expression
\begin{lstlisting}[language=lean]
``(add nat.one nat.one)
\end{lstlisting}
elaborates to \lean{add.{0} nat nat.has_add nat.one nat.one}.
The notation \lean{``(...)} instructs Lean's parser
to interpret the quoted text as a term of type \lean{pexpr}.
Pre-expressions share the same structure as expressions.

Mathematica expressions are analogous to pre-expressions.
They may be type-ambiguous and contain less information than their Lean counterparts.
Thus we normally expect to interpret terms of type \lean{mmexpr} as pre-expressions,
and to use the Lean elaborator to turn them into full expressions.
However, in rare cases an \lean{mmexpr} may already correspond to a full expression.
The unmodified representation of a Lean expression, sent back into Lean,
should be interpreted as the original expression.
We provide two extensible translation functions,
\lean{expr_of_mmexpr} and \lean{pexpr_of_mmexpr},
to handle both of these cases.
Since the implementations are similar we focus our attention on the latter.

The function
\begin{lstlisting}[language=lean]
pexpr_of_mmexpr : trans_env → mmexpr → tactic pexpr
\end{lstlisting}
takes a translation environment and an \lean{mmexpr}, and, using the attribute manager, attempts to return a pre-expression. (Since the tactic monad includes failure, the process may also fail if no interpretation is found.) Interpreting strings as pre-expressions, or, indeed, as expressions, is straightforward. Since Mathematica
integers may be used to represent numerals in many different Lean types,
expressions built with the \lean{mint} constructor are interpreted as untyped numeral pre-expressions.

The \lean{sym} and \lean{app} cases are more complex,
since this part of the translation procedure is extensible by the user.
We define three classes of translation rules:
\begin{itemize}
\item A sym-to-pexpr rule, of type \lean{string × pexpr},
identifies a particular Mathematica symbol with a particular pre-expression.
For example, the rule \lean{("Real",} \lean{``(real))}
instructs the translation to replace the Mathematica symbol \mm{Real}
with the Lean pre-expression \lean{const "real"}.

\item A keyed app-to-pexpr rule is of type
\begin{lstlisting}[language=lean]
string × (trans_env → list mmexpr → tactic pexpr).
\end{lstlisting}
When the procedure encounters an \lean{mmexpr} of the form
\lean{app (sym head) args}---that is,
the Mathematica head symbol \mm{head}
applied to a list of arguments \mm{args}---it will try
to apply all rules that are keyed to the string \lean{head}.
The rules for interpreting arithmetic expressions follow this pattern.
For instance, a rule keyed to the string \lean{"Plus"}
will interpret \mm{Plus[t\textsubscript{1}, ..., t\textsubscript{\emph{n}}]}
by folding applications of \lean{add}
over the translations of \mm{t\textsubscript{1}} through \mm{t\textsubscript{\emph{n}}}.

\item An unkeyed app-to-pexpr rule is of type \\
\lean{trans_env → mmexpr → list mmexpr → tactic pexpr}.
If the head of the application is a compound expression,
or if no keyed rules execute successfully,
the translation procedure will try unkeyed rules.
One such rule attempts to translate the head symbol and arguments independently,
and fold application over these translations.
Another removes instances of the symbol \mm{Hold},
which blocks evaluation of sequences of expressions.
The Lean translation of \mm{Plus[Hold[x, y, z]]}
should reduce to the translation of \mm{Plus[x, y, z]},
but since \mm{Hold[x, y, z]} translates to a sequence of expressions,
this does not match either of the previous rule types.
\end{itemize}

Rules of these three types can be declared by the user
and tagged with the corresponding attribute.
The translation procedure uses Lean's caching attribute manager
to collect relevant rules at runtime.
The mechanism for extending the translation procedure
is thus integrated into theory development.
Translation rules are first-class members of mathematical libraries,
and any project importing a library will automatically have access to its translation rules.

Returning to our example, we have translated the expression \lean{x^2 - 2x + 1} and factored the result, to produce
\mm{Power[Plus[-1, X], 2]}. This is reflected as the Lean \lean{mmexpr}
\begin{lstlisting}[language=lean]
app (sym "Power") [app (sym "Plus") [mint -1, X], mint 2]
\end{lstlisting}
where again:
\begin{lstlisting}[language=lean]
X := app (sym "LeanLocal")
       [str "17.27", str "x", str "bi",
        app (sym "LeanConst") [str "real", []]]
\end{lstlisting}

Applying \lean{pexpr_of_mmexpr} produces the pre-expression
\lean{pow_nat (add (neg one) x) (bit0 one)},
which elaborates to the expression:
\begin{lstlisting}[language=lean]
pow_nat real real_has_pow_nat (add real real_has_add (neg real real_has_neg (one real real_has_one) x) (bit0 nat nat_has_add one nat nat_has_one) : real
\end{lstlisting}

Formatted with standard notation and implicit arguments hidden, we have constructed the term
\begin{lstlisting}[language=lean]
x : real ⊢ (x + -1)^2 : real
\end{lstlisting}
as desired.

\subsection{Translating binding expressions}
Lean's expression structure uses anonymous bound variables to implement its \lean{pi}, \lean{lam}, and \lean{elet} binder constructs. Mathematica, in contrast, has no privileged notion of a binder. The Lean pre-expression
$\lambda$ \lean{x, x + x}
is analogous to the Mathematica expression \mm{Function[x, x+x]}, but the underlying representation of the latter is an application of the \mm{Function} head symbol to two arguments, the symbol \mm{x} and the application expression \mm{Plus[x, x]}. Structurally it is no different from \mm{List[x, x+x]}.

To properly interpret binder expressions,
both translation routines need a notion of an environment.
We extend the Mathematica function \mm{LeanForm} with another argument,
a list of symbols \mm{env} tracking binder depth.
When the translation routine encounters a binding expression,
it creates a new symbol,
prepends it to the \mm{env},
and translates the binder body under this extended environment.
A bound variable \mm{LeanVar[i]} is interpreted as the $i$th entry in \mm{env}.

In the opposite translation direction,
a translation environment is
a map from strings (names of symbols) to expressions,
that is, \lean{trans_env := rb_map string expr}.
The \lean{rb_map} type implements such a map as a red--black tree.
When translating a Mathematica expression such as \mm{Function[x, x+x]},
the procedure extends the environment
by mapping \mm{x} to a placeholder variable,
translates the body under this extended environment,
and then abstracts over the placeholder.
Unlike in Lean,
where \lean{pi}, \lean{lam}, and \lean{elet} expressions
are the only expressions that encode binders,
there are many Mathematica head symbols
(e.g. \mm{Function}, \mm{Integrate}, \mm{Sum})
that must be translated this way.

\section{Querying Mathematica from Lean}
\label{section:mmfromlean}

The translation described in Section~\ref{section:translation} is bidirectional.
Syntax from either system can be embedded and interpreted in the other.
In this section,
we describe the interface used for querying Mathematica from a Lean session,
along with a number of examples of how this interface is used.

\subsection{Connection interface}
\label{subsection:mmfromlean:interface}

Because of the cost of launching a new Mathematica kernel,
it is undesirable to do so every time Lean makes a query.
Instead, we implement a simple server in Mathematica,
which receives requests containing expressions
and returns the results of evaluating these expressions.
Lean communicates with this server by calling a Python client script
via its command line IO interface.
This short script is the only part of the link
that is implemented neither in Lean nor in Mathematica.

This architecture ensures that a single Mathematica kernel will be used for as long as possible, across multiple tactic executions and possibly even multiple Lean projects. To preserve an illusion of ``statelessness,'' each Mathematica evaluation occurs in a new context which is immediately cleared. While this avoids accidental leaks of information, it is not a watertight seal, and users who consciously wish to preserve information between sessions can do so.

The translation procedure is exposed in Lean using the tactic framework via
\begin{lstlisting}[language=lean]
meta def mathematica.execute : string → tactic mmexpr
\end{lstlisting}
This tactic evaluates the input string in Mathematica
and returns a term with type \lean{mmexpr} representing the result of the computation.
From this basic tactic, it is easy to define variants, e.g.:
\begin{lstlisting}[language=lean]
run_command_using : (string → string) → expr → string → tactic pexpr
\end{lstlisting}
The first argument is a Mathematica command, including a placeholder bound variable, which is replaced by the Mathematica representation of the \lean{expr} argument. The \lean{string} argument is the path to a file which contains auxiliary definitions, usable in the command. This variant will  apply the back-translation \lean{pexpr_of_mmexpr} to produce a \lean{pexpr}.

Another variant, \lean{execute_global : string → tactic mmexpr}, evaluates its input in Mathematica's global context.

Going back to our running example from Section \ref{section:translation}, assuming \lean{e} is the unfactored expression, we would call
\begin{lstlisting}[language=lean]
run_command_on (λ s, s ++ " // LeanConvert // Activate // Factor") e
\end{lstlisting}
to produce a pre-expression representing the factored form of \lean{e}. (Recall that the Mathematica syntax \mm{x // f} reduces to \mm{f[x]}.)
In fact, we can define
\begin{lstlisting}[language=lean]
meta def factor (e : expr) : tactic pexpr :=
run_command_on (λ s, s ++ " // LeanConvert // Activate // Factor") e
\end{lstlisting}
or a variant that elaborates the result into an \lean{expr} with the same type as \lean{e}.

\subsection{Verified interaction}
\label{subsection:mmfromlean:verified}

So far we have described how to
embed a Lean expression in Mathematica, manipulate it,
and import the result back into Lean.
At this point, the imported result is simply a new expression:
no connection has been established between the original and the result.
In our factoring example,
we expect the two expressions to be equal.
If we were computing an antiderivative,
we would expect the derivative of the result to be equal to the original input.
More complex return types can lead to more complex relations.
For example, an algorithm using Mathematica's linear programming tools
to verify the unsatisfiability of a system of equations
may return a certificate that must be converted into a proof of falsity.

Users may simply decide to trust the translation and CAS computation
and assert without proof that the result has an expected property.
An example using this approach is given in Section~\ref{subsection:mmfromlean:unverified}.
Of course, the level of trust needed to do this is unacceptably high for many situations.
We are often interested in performing \emph{certifiable} calculations in Mathematica,
and using the certificates to construct proofs in Lean.

It would be hopeless to expect one tool to verify all results.
Rather, for each common computation,
we will have a tactic script
that attempts to prove the appropriate relation between input and output.
``Uncommon'' or one-off computations can be verified in-line by the user.
This method of separating search (or computation) and verification
is discussed at length by Harrison and Th\'ery \cite{Harrison1998} and by many others.
It turns out that a surprising number of algorithms
are able to generate certificates to this end.

The tactics used in this section, along with more examples,
are available in the supplementary information to this paper.
These examples are not meant to be exhaustive,
but rather to illustrate the ease with which Mathematica can be accessed:
each is fairly simple to implement.

\paragraph{Factoring.}
\label{subsection:verification:factoring}
In our running example, we have used Mathematica to construct the Lean expression \lean{(x + -1)^2 : real}. We expect to find a proof that \lean{x^2 - 2*x + 1 = (x + -1)^2}.
This type of proof is easy to automate with Lean's ring normalization tactic:

\begin{lstlisting}[language=lean]
meta def eq_by_ring (l r : expr) : tactic expr :=
do gl ← mk_app `eq [l, r],
   mk_inhabitant_using gl ring
     <|> fail "unable to simplify"
\end{lstlisting}

Using this machinery, we can easily write a tactic \lean{factor}
that, given a polynomial expression,
factors it and adds a constant to the local context asserting equality.
(The theorem \lean{pow_two_nonneg} proves that the square of a real number is nonnegative.)

\begin{lstlisting}[language=lean]
example (x : ℝ) : x^2-2*x+1 ≥ 0 :=
by factor x^2-2*x+1 using q; rewrite q; apply pow_two_nonneg
\end{lstlisting}

We provide more examples of this tactic in action in the supplementary material,
including one that factors \lean{x^10-y^10}:
\begin{lstlisting}[language=lean]
(x + -1 * y) * (x + y) * (x^4 + -1 * x^3 * y + x^2 * y^2 + -1 * x * y^3 + y^4) * (x^4 + x^3 * y + x^2 * y^2 + x * y^3 + y^4)
\end{lstlisting}

In general, factoring problems are easily handled by this type of approach,
since the results serve as their own certificates.
Factoring integers is a simple example of this:
to verify, simply multiply out the prime factors.
Dually, primality certificates can be generated and checked \cite{Pratt1975}.

Factoring matrices is slightly more complex. Mathematica implements a number of common matrix decomposition methods, whose computation can be verified in Lean by re-multiplying the factors. We can use these tools to, e.g., define a tactic \lean{lu_decomp} which computes and verifies the LU decomposition of a matrix.
\begin{lstlisting}[language=lean]
example : ∃ l u, is_lower_triangular l
   ∧ is_upper_triangular u
   ∧ l ** u = [[1, 2, 3], [1, 4, 9], [1, 8, 27]] :=
by lu_decomp
\end{lstlisting}

\paragraph{Solving polynomials.}

Mathematica implements numerous decision procedures and heuristics
for solving systems of equations.
Many of these are bundled into its \mm{Solve} function.
Over some domains,
it is possible to verify solutions in Lean using the simplifier,
arithmetic normalization, or other automation.
Lean's \lean{norm_num} tactic,
which reduces arithmetic comparisons between numerals,
is well suited to verifying solutions to systems of polynomial equations.
The tactic \lean{solve_polys} uses \mm{Solve} and \lean{norm_num} to prove such theorems:
\begin{lstlisting}[language=lean]
example :  ∃ x y : ℝ, 99/20*y^2 - x^2*y + x*y = 0
  ∧ 2*y^3 - 2*x^2*y^2 - 2*x^3 + 6381/4 = 0 :=
by solve_polys
\end{lstlisting}

Users familiar with Mathematica may recall that \mm{Solve} outputs a list of lists of applications of the \mm{Rule} symbol, each mapping a variable to a value. Each list of rules represents one solution. A \mm{Rule} has no general correspondent in Lean, and it would involve some contortion to translate this output and extract a single solution in the proof assistant. However, it is easy to perform this transformation within Mathematica, and processing the result of \mm{Solve} \emph{before} transporting it back to Lean makes the procedure much simpler to implement. This type of consideration appears often: some transformations are more easily achieved in one system or the other.

%
%
\paragraph{Linear arithmetic.}
Many proof assistants provide tools for automatically proving linear arithmetic goals,
or equivalently for proving the unsatisfiability of a set of linear hypotheses.
There are various approaches to this,
including building proof terms incrementally
using Fourier--Motzkin elimination~\cite{Williams1986}.
Alternatively, linear programming techniques can be used to generate certificates of unsatisfiability.
A certificate for the unsatisfiability of
$\{p_i(\bar x) \leq 0 : 0 \leq i \leq n \}$
is a solution to the dual program,
that is, a list of rational coefficients $\{c_i : 0 \leq i \leq n\}$ such that
$\sum_{0\leq i \leq n} c_i \cdot p_i = q > 0$
for some constant polynomial $q$.
Equivalently, this list serves as a witness for Farkas' lemma~\cite{schrijver:86}.

The tactic \lean{linarith} implemented in Lean's \mathlib is in effect
a generic certificate checker for linear arithmetic.
Given a function that implements a simple certificate-finding interface,
\lean{linarith} will preprocess the context, retrieve a certificate,
and convert the certificate into a proof of the goal.
Unlike Coq's similar \lean{lra}~\cite{besson:07},
which reflexively calls a proven-correct certificate checker,
\lean{linarith} constructs a proof term from the certificate using ring normalization.

By default, \lean{linarith} uses an unverified Fourier--Motzkin solver implemented in Lean
to produce certificates.
But this solver is a black box to the rest of the tactic.
It can be desirable on large problems to
use a more efficient algorithm, for instance one based on the simplex method,
and \lean{linarith} provides a configuration option to change the search function.

We provide an alternate certificate oracle that uses Mathematica's linear programming functionality.
It takes only 20 lines of Lean code to implement a full drop-in replacement
for the Fourier--Motzkin module.
Most of this is string processing to create the appropriate call to Mathematica;
the certificate search is summarized by

\begin{verbatim}
L = Part[#, 2]& /@
  FindInstance[{cstrs}&&{nngs}&&{pos}, {vars}, Rationals][[1]];
(LCM@@Denominator@L)*L
\end{verbatim}
where \verb|{cstrs}&&{nngs}&&{pos}| is a system of linear constraints in variables \verb|{vars}|
describing an appropriate solution to the dual problem.
The final line guarantees a solution in the nonnegative integers.
This function produces a (Mathematica) list of integers,
which our link reinterprets in Lean as a list of natural numbers
as required by the \lean{linarith} oracle specification.

While communication overhead makes this approach slightly slower
than the native Fourier--Motzkin solver on small problems,
the Mathematica oracle succeeds on every \lean{linarith} test case, e.g.:
\begin{lstlisting}[language=lean]
example (x y z : ℚ) (h1 : 2*x  < 3*y) (h2 : -4*x + 2*z < 0)
  (h3 : 12*y - 4* z < 0) : false :=
by linarith {oracle := mm_oracle}
\end{lstlisting}

\subsection{Unverified interaction}
\label{subsection:mmfromlean:unverified}

The applications of CAS to interactive proving go beyond verified computations.
We emphasize the word ``interactive'' in the name of the field:
proof assistant users may want to query the CAS for guidance
as they work on a proof on their own.
This requires a reasonable expectation of correctness from the CAS
but no formal verification.
Other more credulous users may be willing to take the CAS at its word
and trust its output.

In this section we explain a number of ways that
the results of Mathematica computations can be used from within Lean
without verification.

\paragraph{Error checking.}

Mathematica's \mm{FindInstance} function
is a frontend for a collection of different solvers
that try to find variable instantiations that satisfy a given predicate.
This function can be used from Lean
to check whether a proof goal is in fact provable.
We define a tactic \lean{plausibility_check}
which fails if Mathematica is able to find a variable assignment
that satisfies the local hypotheses and the negation of the current goal.
This tactic is similar to a very lightweight version of Isabelle's Nitpick \cite{blanchette2010}.
The first example below fails when Mathematica decides that the goal does not follow;
the second succeeds.
\begin{lstlisting}[language=lean]
example (x : ℝ) (h1 : sin x = 0) (h2 : cos x > 0) : x = 0 :=
by plausibility_check; admit

example (x : ℝ) (h1 : sin x = 0) (h2 : cos x > 0)
        (h3 : -pi < x ∧ x < pi) : x = 0 :=
by plausibility_check; admit
\end{lstlisting}

\paragraph{Notebook-style interaction.}

The standard mode of interaction with Mathematica and other CAS tools
is through a notebook interface,
similar to a read--eval--print loop (REPL)
with extra functionality for back references and displaying output.
This style encourages using the CAS as a tool for exploration,
since the user can interactively change and add to the system state
and visualize the output of entries.
While some proof assistants do implement REPLs
for constructing proofs via tactic application
or inspecting declarations in an environment,
the notion of evaluation is much more limited in a proof assistant than in a CAS,
limiting the ways in which one can use such an interface.
Notebooks that support proof assistant languages, such as Observable\footnote{\url{https://observablehq.com/@bryangingechen/hello-lean-prover}},
tend to be used only for small demonstrations.

The appeal of notebook-style interaction with a proof assistant grows
with access to the evaluation and visualization tools of a CAS.
We have implemented a top-level Mathematica code block command in Lean
that approximates this kind of interface (Figure~\ref{figure:mm-block}).
Inside a block of Mathematica code embedded in a Lean file,
the user can write arbitrary Mathematica commands,
inserting antiquoted Lean expressions at any point.
Upon evaluation, these antiquoted expressions are translated to Mathematica expressions.
The result of the evaluation is translated back to a Lean expression
and displayed in the editor infoview.

\begin{figure}
\begin{lstlisting}[language=lean]
begin_mm_block

"Solve[Sin[x] == 0 && 2 < x < 4, x, Reals][[1]][[1]][[2]]";

"Factor["(x^2-2*x+1)"]";

end_mm_block
\end{lstlisting}

\caption{
    This code block can be inserted at any top-level position
    in a Lean source file.
    It will evaluate the two Mathematica commands successively,
    and display the output of each in the Lean editor.
    The input to \mm{Factor} is an antiquoted Lean expression,
    and the displayed output is also a Lean expression.
}

\label{figure:mm-block}
\end{figure}

\begin{figure}






\includegraphics[width=\textwidth]{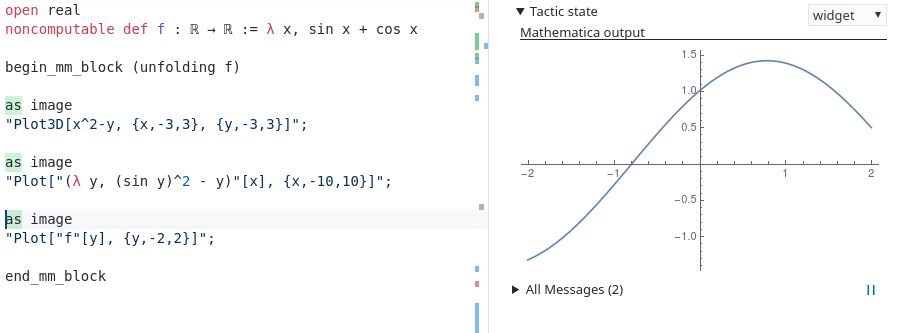}

    \caption{
        This embedded Mathematica code block generates three images.
        The first is given by a pure Mathematica command.
        The second and third plot antiquoted Lean functions.
        The parameter \lean{(unfolding f)} at the beginning of the block
        allows Mathematica to ``see through" the definition of \lean{f}.
        The right hand pane of the editor displays the third plot
        when the cursor hovers over that line.
    }

    \label{figure:mm-block-img}
    \end{figure}

One of Mathematica's greatest strengths is
its toolbox for generating plots, graphs, and images.
Our code embedding makes these available from within Lean (Figure~\ref{figure:mm-block-img}).
Prefixing the Mathematica command with an annotation \lean{as image}
requests that the output be displayed graphically instead of textually.
Users can access the full range of Mathematica's visualization tools
to plot and inspect Lean terms.

The image display makes use of Ayers' \emph{widget} feature for
Lean\footnote{\url{https://github.com/leanprover-community/lean/blob/master/library/init/meta/widget/basic.lean}}
which allows metaprograms to display arbitrary HTML in the editor.
The syntactical oddities of embedding Mathematica code
reflect limitations of the Lean 3 parser.
Lean 4, under development at the time of submission of this paper,
features a highly customizable parser
with precisely this kind of domain specific language embedding in mind~\cite{Ullrich2020}.


\paragraph{Axiomatized computations.}

Since it is possible to declare axioms from within the Lean tactic framework, we can axiomatize the results of Mathematica computations dynamically. This allows us to access a wealth of information within Mathematica, at least when we are not concerned about complete verification. One interesting application is to query Mathematica for special function identities. While these identities may be difficult to formally prove, trusting Mathematica allows us to find some middle ground. The \lean{prove_by_full_simp} tactic uses Mathematica's \mm{FullSimplify} function to reduce the Bessel function expression on the left, and after checking that it is equal to the one on the right, adds this equality as an axiom in Lean:
\begin{lstlisting}[language=lean]
example : ∀ x, x*BesselJ 2 x + x*BesselJ 0 x = 2*BesselJ 1 x :=
by prove_by_full_simp
\end{lstlisting}

We can also define a tactic that uses Mathematica to obtain numerical approximations of constants, and axiomatizes bounds on their accuracy:
\begin{lstlisting}[language=lean]
approx (100 * BesselJ 2 (13 / 25)) (0.001 : ℝ)
\end{lstlisting}
declares an axiom stating that
\begin{lstlisting}[language=lean]
75977 / 23000 < 100 * BesselJ 2 (13 / 25)
  ∧ 100 * BesselJ 2 (13 / 25) < 76023 / 23000.
\end{lstlisting}

\section{Querying Lean from Mathematica}
\label{section:leanfrommm}
The use of computer algebra in mathematics
is largely limited to exploration and discovery.
Finished proofs often avoid using these tools to justify claims
or even fail to mention them entirely.
Outside of a few very specific domains,
systems like Mathematica have no internal notion of proof or correctness.
There are many documented instances of
bugs and unexpected behavior in computer algebra systems \cite{bailey2006},
making concerns about this black-box nature more than just theoretical.
Even the semantics for certain computations can be vague:
reducing \mm{(x\^{}2 - 1)/(x - 1)} to \mm{x + 1} is correct when considered as polynomial division,
but computer algebra systems use this same notation to refer to a function of \mm{x}.

Integrating a proof system into a mature CAS such as Mathematica is an enormous engineering task.
A more realistic approach is
to use a translation procedure to ``borrow'' a proof language and semantics from a proof assistant
on translatable domains.
A proposition relating the input and output of a CAS evaluation
can be exported to and proved in the proof assistant,
which can return a proof term.
This is morally similar to the ad hoc verification described in the previous section.
While no general guarantee is claimed, individual computations can be checked.

More generally, the exploratory uses of CAS tools rely on databases of definitions and examples.
Mathematica features enormous data sets, mathematical and otherwise,
but these mainly describe computational objects,
with only a few examples from pure mathematics~\cite{Ford17}.
Connecting Lean to Mathematica allows the Lean library \mathlib
to serve as a collection of definitions, theorems, and proofs
that the CAS can inspect and process.

\subsection{Connection interface}
\label{subsection:leanfrommm:interface}

We use Mathematica's external command interface \mm{StartProcess}
to establish a connection to Lean from a Mathematica notebook.
The notebook communicates with a Lean server session
via the same interface used by Lean editors.
All calls to Lean are directed to a particular server process,
which allows for the option to run and query multiple server processes at once.
The standard mode of interaction, though,
is to begin a session by defining a \mm{ProcessObject} expression
\begin{verbatim}
Lean = LeanMode[]
\end{verbatim}
and to send all requests to this single \mm{ProcessObject}.
As in Section \ref{subsection:mmfromlean:interface},
this server--client interface allows us to preserve Lean environment state as needed
and to avoid the cost of launching a new process for each query.


A low level communication function
\begin{verbatim}
SendToLeanServer[p_ProcessObject, content_String]
\end{verbatim}
evaluates an arbitrary server request \mm{content} in Lean process \mm{p}.
We build higher level tools on top of this.
For instance, a function
\begin{verbatim}
RunLeanTactic[p_ProcessObject, x_, t_String]
\end{verbatim}
takes an arbitrary Mathematica expression \mm{x}
and a string \mm{t} naming a Lean tactic,
and returns the result of calling \mm{t} on the Lean interpretation of \mm{x}.
It does so by constructing the top-level syntax needed to perform this operation
and passing this syntax to \mm{RunLeanTactic}.




\subsection{Applications}
\label{subsection:leanfrommm:applications}

\begin{figure}
  \includegraphics[width=\linewidth]{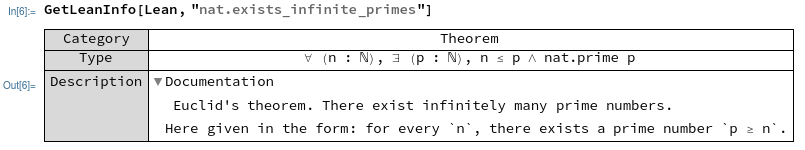}
  \caption{Retrieving the declaration information from \lean{nat.exists_infinite_primes}.}
  \label{figure:get-lean-info}
\end{figure}

\paragraph{Querying about certain declarations.}
Mathematica contains many databases
ranging over a huge variety of topics.
A major motivation to connect Lean to Mathematica
is to treat the proof assistant library as another such database.
As a simple example of this kind of use,
we implement a function
\begin{verbatim}
GetLeanInfo[p_ProcessObject, decl_String]
\end{verbatim}
that displays information about the declaration named \mm{decl}
found in the Lean process \mm{p} (Figure~\ref{figure:get-lean-info}).
This information includes the declaration's category
(is it an axiom, a definition, a theorem?), its type,
and any documentation associated with it.
The natural language description in this output
is taken from the same source as the \mathlib API documentation~\cite{Door20}.

By default, \mm{GetLeanInfo} returns a structure whose fields are strings,
as this is most convenient to print and display.
But it is very easy to instead retrieve the type as an expression.
While an arbitrary \mathlib declaration is unlikely to fully translate
to a Mathematica counterpart,
one may wish to perform further processing on the syntax tree of the type.
It is also a simple matter to retrieve the body of the declaration,
as we do in some examples below.

\paragraph{Displaying propositional proofs.}
\begin{wrapfigure}{L}{.4\linewidth}
  \includegraphics[width=\linewidth]{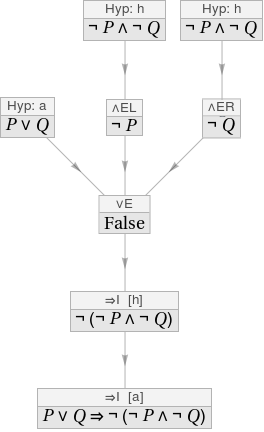}
  \caption{A natural deduction diagram generated from a Lean proof term.}
  \label{figure:nd}
\end{wrapfigure}
Mathematica's \mm{TautologyQ} and \mm{FullSimplify} functions
serve as complete SAT solvers.
However, both are black boxes, in that they produce no certificate or justification.
Indeed, the system has no established proof language for propositional logic.
On the other hand,
Lean comes equipped with a number of proof-producing decision procedures for this domain.
For this example, we use \lean{intuit},
as it uses a small grammar of proof rules.

We define a minimal propositional proof calculus in Mathematica
that mirrors the calculus in Lean.
That is, we introduce head symbols
\mm{AndIntro}, \mm{OrIntroLeft}, \mm{FalseElim}, etc.,
and add \mm{LeanForm} translation rules
that map Lean's \lean{and.intro}, \lean{or.inl}, \lean{false.elim}, etc.\
to their corresponding symbols.
We can then state a propositional theorem in Mathematica,
prove it in Lean,
and interpret the resulting proof term in our calculus.
While it would certainly be possible
to implement the Lean proof search procedure in Mathematica directly,
this approach ensures that the proof is correct,
as it has been checked by Lean.

We emphasize here that the input formula is a pure Mathematica object,
for example \mm{Implies[Or[P, Q], Not[And[Not[P], Not[Q]]]]}.
Generating these proofs does not require any knowledge
of Lean syntax or the encoding of Mathematica syntax in Lean.

The resulting Mathematica proof object can be computed with in any number of ways.
We implement a function which displays the proof as a natural deduction diagram
(Figure~\ref{figure:nd}).
There is no fundamental reason why this approach cannot be extended to richer logics
such as first-order logic.
The difficulties lie in representing a calculus for these logics in Mathematica
and generating proofs in Lean that can be translated to such a calculus.
(Many proof tools in Lean use higher-order constructs that may be difficult to directly translate.)

\paragraph{Displaying arbitrary proofs.}
Natural de\-duc\-tion-style proofs are most commonly seen
in pure propositional and first-order logic,
making the diagrams of the previous section particularly familiar.
However, the presentation style can be adapted to richer logics.
Lean's \lean{#explode} command, implemented in \mathlib,
formats a Lean proof term in a way resembling a Fitch natural deduction diagram.

Natural deduction proofs are not known for their brevity.
The textual output of \lean{#explode}, viewed in a Lean editor session,
can be overwhelmingly long.
It is much more enlightening to be able to expand and fold the output.
We implement a function \mm{GetLeanProof[p\_ProcessObject, decl\_String]} in Mathematica
that retrieves the exploded output of the Lean declaration named \mm{decl}
from the Lean process \mm{p}
and formats it as an interactive object in the Mathematica notebook.
Users can drill down into the details of the proof
as deep as needed to understand its structure.

\begin{figure}
  \centering
  \includegraphics[width=\textwidth]{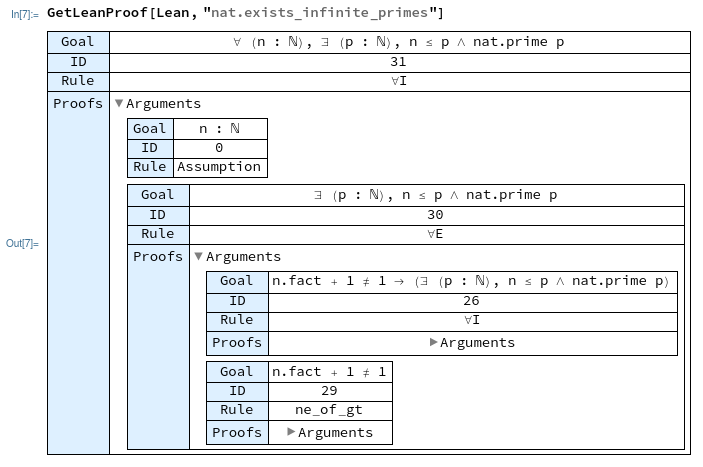}
  \caption{An expandable Fitch-style diagram generated from a Lean proof term.}
  \label{figure:explode}
\end{figure}

Figure~\ref{figure:explode} displays  the beginning of an exploded proof
of the fact that there are infinitely many primes.
At each node of the diagram there appears a goal,
labeled with a unique index.
The goal may be referenced by this index in subsequent nodes.
Each node is justified by the application of a particular rule:
the original goal is proved by an application of universal introduction,
after which we introduce a fresh variable \lean{n} by the assumption rule.
Some nodes are justified by the application of library lemmas,
for instance node 29,
in which the goal \lean{n! + 1 ≠ 1} is justified
by applying the lemma \lean{ne_of_gt}.
Depend on the rule applied,
a node may have arguments,
represented as nodes themselves.
An application of universal introduction takes two arguments,
the newly introduced variable and a proof of the remaining goal;
introducing this variable is an atomic step
with no arguments.

Due to the Curry--Howard correspondence,
the same technique can be used to inspect data-valued expressions,
e.g.\ to unfold parts of a declaration defining a natural number.

This application demonstrates how mathematicians can use our link
to explore the proof assistant library without leaving the CAS.
It takes advantage of Mathematica's highly developed user interface
to visualize the details of arguments
in a way that is nearly impossible to do from within Lean,
especially for proofs that use substantial automation
or that do not have easily legible proof scripts.

\paragraph{Proving Mathematica goals.}
As with the examples from propositional logic,
the Wolfram Language is rich enough to express some propositions
that can be proved in Lean.
We implement a function
\begin{verbatim}
ProveUsingLeanTactic[p_ProcessObject, x_,t_String]
\end{verbatim}
that takes an arbitrary Mathematica expression \mm{x}
and a Lean tactic string \mm{t}
and tries to translate and prove the Mathematica expression,
finally displaying the exploded proof.
While in principle this accepts any tactic proof,
it is not convenient to interactively write a proof from the Mathematica notebook,
and so this is typically used for statements
that can be solved with one or two lines.

As mentioned at the beginning of this section,
we see this as informally lending the semantics of Lean
to Mathematica expressions.
The semantic value of the proposition \mm{x} in Mathematica
is merely if and whether it simplifies to \mm{True} or \mm{False}.
With our translation and link,
we provide a true proof language for some subset of these expressions.

\paragraph{Displaying significant proof steps.}
The full output of \lean{#explode} may sometimes contain too much detail,
as the logical steps like universal introduction may be unenlightening.
An alternate way to visualize a proof is
to identify and display important lemmas or relevant steps used within.
Lean's simplifier,
heuristic quantifier instantiation procedure,
and other general-purpose proof tactics
search for lemmas in the Lean library to solve a goal.
It is possible to inspect the proof terms generated by these tactics
and extract theory lemmas, or in some cases,
to implement versions of these tactics that produce a list of lemmas used.
The types of the instances of these lemmas appearing in a proof term
can be interpreted in Mathematica and displayed.
Finding all and only the ``interesting'' lemmas
is a difficult and poorly specified problem,
but it is reasonable to implement a first-pass heuristic.

As an example, we do so in the context of set normalization.
Mathematica has no built-in handling for arbitrary sets,
but proofs of propositions
such as $A \cap (B \cup \bar A) = A \cap B$
are easily found with Lean's simplifier.
Noting that the relevant lemmas used by \lean{simp} state that
\lean{A ∩ (B ∪ -A)} \lstinline[language=lean]{=} \lean{(A ∩ B) ∪ (A ∩ -A)},
\lean{A ∩ -A} \lstinline[language=lean]{=} \lstinline[language=lean]{∅},
and \lean{(A ∩ B) ∪ ∅} \lstinline[language=lean]{=} \lean{A ∩ B},
we can return these lemmas to Mathematica and display them as a ``proof sketch.''
Note that there is no need to add translation rules for these lemmas themselves;
alignments between the constants for
union, intersection, complement, and equality
are enough.
This limits the need for a long list of translations
and makes the procedure relatively robust to the introduction of new simplifier rules.

A similar application involves the use of a relevance filtering algorithm.
Given a target expression,
such an algorithm will return a list of declarations in the environment
that, heuristically, may be useful to prove the target.
Both symbolic and probabilistic relevance filters
have been implemented in other systems
and are used for lemma selection for tools such as Isabelle's Sledgehammer \cite{blanchette2016}.
We have implemented a rudimentary relevance filter in Lean.
Using this tool,
one can state a conjecture in Mathematica
and receive a list of facts that may be of use to prove it,
without depending on automation in Lean to actually find a proof.



\section{Concluding thoughts}
\label{section:conclusion}

\subsection{Related work}
\label{subsection:conclusion:related}
The following discussion is not meant to be comprehensive,
but rather to indicate the many ways
in which one can approach connecting ITP and computer algebra.

Harrison and Th\'ery \cite{Harrison1998}
describe a ``skeptical'' link between HOL and Maple
that follows a similar approach to our bridge.
Computation is done in a standard, standalone version of the CAS
and sent to the proof assistant for certification.
The running examples used are factorization of polynomials and antiderivation.
The discussion is accompanied by an illuminating comparison
of proof search to proof checking,
and the relation to the class NP.
Delahaye and Mayero \cite{Delahaye2005}
provide a similar link between Coq and Maple,
specialized to proving field identities.
Both projects tackle only the scenario in which the proof assistant drives the CAS.

This skeptical approach is also taken in some projects
that require one specific type of computer algebra computation
instead of a generic link.
Harrison~\cite{harrison:15} computes Wilf--Zeilberger certificates in Maxima
and verifies them in HOL Light.
While this work does establish an interface between the two systems,
Harrison notes the convenience of a ``manual'' version,
where users generate certificates in Maxima
and transfer them to HOL Light by hand.
Chyzak, Mahboubi, Sibut-Pinote, and Tassi~\cite{chyzak:14,mahboubi:19}
use certificates from Maple in the verification of a critical lemma
for proving the irrationality of $\zeta(3)$.
Here the CAS results are only transferred manually.
These instances of manual translation
between a CAS notebook and a proof assistant
suggest future work on an integrated user interface,
described in the final section of this paper.

Ballarin and Paulson \cite{Ballarin1999}
provide a connection between Isabelle
and the computer algebra library $\Sigma^\text{IT}$ \cite{bronstein1996}
that is more trusting than the previous skeptical approach.
They distinguish between sound and unsound algorithms in computer algebra:
roughly, a sound algorithm is one whose correctness is provable,
while an unsound algorithm may make unreasonable assumptions about the input data.
Their link accepts sound algorithms in the CAS as oracles.
A similarly trustful link between Isabelle and Maple,
by Ballarin, Homann, and Calmet \cite{Ballarin1995},
allows the Isabelle user to introduce equalities derived in the CAS as rewrite rules.
A third example by Seddiki, Dunchev, Khan-Afshar, and Tahar \cite{Seddiki2015}
connects HOL Light to Mathematica via OpenMath,
introducing results from the CAS as HOL axioms.

A related, more skeptical, approach is
to formally verify CAS algorithms
and incorporate them into a proof assistant via reflection.
This approach is taken by D\'en\`es, M\"ortberg, and Siles \cite{Denes2012},
whose CoqEAL library implements a number of algorithms in Coq.

Kerber, Kohlhase, and Sorge \cite{kerber:1998}
describe how computer algebra can be used in proof assistants
for the purpose of proof planning.
They implement a minimal CAS
which is able to produce high-level sketch information.
This sketch can be processed into a proof plan,
which can be further expanded into a detailed proof.

Alternatively, one can build a CAS inside a proof assistant without reflection,
such that proof terms are carried through the computation.
Kaliszyk and Wiedijk \cite{Kaliszyk2007}
implement such a system in HOL Light,
exhibiting techniques for simplification, numeric approximation, and antiderivation.

Going in the opposite direction,
CAS users may want to access ATP or ITP systems.
Adams et al. \cite{Adams2001}
use PVS to verify side conditions generated in computations in Maple;
Gottliebsen, Kelsey, and Martin \cite{gottliebsen:05} make use of similar ideas.
Systems such as Analytica \cite{Bauer1998} and Theorema \cite{Buchberger2016}
provide ATP- or ITP-style behavior from within Mathematica.
Axiom \cite{daly2005} and its related projects
provide a type system for computer algebra,
which is claimed to be ``almost'' strong enough
to make use of the Curry--Howard correspondence.

\subsection{Future work}
\label{subsection:conclusion:future}

There is much room for an improved interface under the current ITP--CAS relationship.
We imagine a link integrated with Lean's supported editors,
where the user can communicate with Mathematica in true notebook style
with access to the environment at a particular point in a Lean file.
In the imagined Mathematica notebook,
quoted Lean pre-expressions will be elaborated
in the environment at the demarcated point in the Lean file,
and then reflected and processed in Mathematica.
The results can then be easily exported to the Lean file.
The notebook is a standard way of interacting with computer algebra systems
and contributes to their utility in exploration and discovery.
While the embedded Mathematica code blocks described in Section~\ref{subsection:mmfromlean:unverified}
are a first-degree approximation to this kind of interaction,
a full-fledged notebook interface is more natural to use.
Our link implements all the translation and communication features needed for such an integration
but substantial UI engineering is needed to make it a reality.

The server interface described in Section \ref{subsection:mmfromlean:interface}
only supports sequential evaluation of Mathematica commands.
Both systems support parallel computation,
and integrating the two could increase the utility of this link for large projects.
Improvements to the foreign function interface in Lean 4
may allow the physical connection to be made more robust.

With the exception of the server running in Mathematica,
the components of this link
are generally adaptable to other computer algebra systems.
More broadly, we see this project as part of a general trend.
The various computer-based tools used in mathematical research,
by and large, are independent of each other.
It requires quite a lot of copying, pasting, and translating to, for example,
compute an expression in Magma~\cite{Bosma1997},
verify its side conditions in Z3~\cite{demoura:bjorner:08},
visualize the results in Mathematica,
and export relevant formulas to \LaTeX.
Unified frameworks have been proposed and implemented \cite{rabe2013} but are not widely used.
Because they provide a strict logical foundation,
precise semantics,
and possibility of verification,
proof assistants are strong candidates to center translation networks between systems.

\begin{acknowledgements}
We acknowledge Jeremy Avigad, Jasmin Blanchette,
Ian Ford, Johannes H\"olzl, Jos\'e Mart\'in-Garc\'ia,
Leonardo de Moura, James Mulnix, Michael Trott, and the Lean community
for help, suggestions, and support.
\end{acknowledgements}

\bibliographystyle{spmpsci}      
\bibliography{citations}   

%
%

\end{document}